\documentclass[onecolumn]{mn2e}
\usepackage{bm, graphicx}
\begin{document}

\title[Thin lens accuracy]{Accuracy of the thin-lens approximation
in strong lensing by smoothly truncated dark matter haloes}

\author[S. Frittelli and T. P. Kling]
       {Simonetta Frittelli$^1$\thanks{frittelli@duq.edu}
       and Thomas P. Kling$^2$\thanks{tkling@bridgew.edu}\\
    $^1$Department of Physics, Duquesne University,
        Pittsburgh, PA 15282\\
        $^2$ Department of Physics, Bridgewater State Unuiversity,
Bridgewater, MA 02325}

\maketitle
\begin{abstract}

The accuracy of mass estimates by gravitational lensing using the
thin-lens approximation applied to Navarro-Frenk-White mass models
with a soft truncation mechanism recently proposed by Baltz,
Marshall and Oguri is studied. The gravitational lens scenario
considered is the case of the inference of lens mass from the
observation of Einstein rings (strong lensing). It is found that the
mass error incurred by the simplifying assumption of thin lenses is
below 0.5\%.  As a byproduct, the optimal tidal radius of the soft
truncation mechanism is found to be at most 10 times the virial
radius of the mass model.

\end{abstract}


\section{Introduction} \label{intro:sec}

In recent years gravitational lensing has become the main tool in
use to estimate the amount and distribution of matter of
astrophysical systems of various kinds and sizes, from single
compact objects to clusters of galaxies.  Some very wide-ranging
predictions rest on the accuracy of such mass estimates, most
notably the value of the dark energy density and, with it, the time
since the big bang. The predictive power of gravitational lensing as
a tool to study individual deep objects also hinges heavily on the
accuracy of the method. Consequently, any studies of the degree of
accuracy of the method are naturally warranted.  In this paper we
describe a study that provides a quantitative measure of the degree
of accuracy of the method in a particular context.  Together with
those of a previous study of ours \cite{kf}, these results are among
the first quantitative estimates of the accuracy of the thin-lens
approximation applied to gravitational lensing systems.

Our conceptualization of a gravitational lensing system is as
follows: There is a physical ``lens'' that gravitationally disturbs
the path of lightrays from a source to the observer.  The
theoretical framework for this situation would be a spacetime with a
metric that solves the Einstein equations with a mass density source
that describes the lens.  The lightrays from the source to the
observer would be null geodesics connecting both points.  Even
though the gravitational lensing event in itself (the null rays) is
very amenable to a straightforward computational strategy, there is
an intrinsic difficulty with this theoretical framework:  the
metric.  This is determined by the mass density of the lens, also
known as the density profile, and this, at this time, is still
subject to modeling.

The conventional approach consists in making use of the thin-lens
approximation to calculate the bending of lightrays.  The
conventional approach also makes use of the density profile assumed
for the physical system.  However, the quantities of interest in the
conventional approach require a set of manipulations on the density
profile that differ significantly from those needed in the
theoretical framework.  One significant difference resides in the
fact that most quantities of interest in the conventional approach
require the mass density integrated along the line of sight, a
quantity referred to as the projected mass.

The conditions that a mass density must meet as a function in space
in order to be useful for the conventional approach are much weaker
than what is needed for the theoretical framework. The conventional
approach admits density profiles that are not integrable over all
space, namely, that have slow fall-off rates (slower than $r^{-3}$).
Admittedly, such density profiles are unphysical in the sense that
they do not represent the isolated systems that they intend to
model. Nevertheless, they continue to be used because they are known
to be a close approximation to the underlying physical system within
a finite region centered on the lens.   The question arises as to
how to ``terminate'' such unphysical density profiles in the region
where the approximation is not valid anymore (i.e., far from the
lens).

The simplest way to terminate an unphysical density profile is by
``hard'' truncation, that is: by setting the profile discontinuously
to zero beyond some radius.  Arguably, this mechanism results in a
density profile with questionable physical sense, especially because
the truncation radius is largely arbitrary but also because
discontinuous truncation introduces significant spurious
magnification of images.  However, the underlying system may be
better approximated by such a drastically truncated density profile
than by a non-integrable one, so the method is not devoid of merit.

Alternatively to hard truncations, various ``soft'' truncation
mechanisms have been proposed, which normally consist of
multiplicative factors that speed up the fall-off rate of the
density profile in order to ensure a finite total mass.

In the present work we investigate how well the conventional
approach to gravitational lensing estimates the mass of a lens,
compared to the theoretical prediction. There are several subtleties
associated with this seemingly straightforward task.  To start with,
the physical system itself (the actual density profile of a cluster
in all space) is not known.  What is known is a good representation
of the density profile within a certain radius, referred to as the
virial radius. Since the ``true'' density profile of the lens must
be integrable over all space,  we take the view that a profile with
a finite total mass is the ``true'' system, and any other profile
with a fall-off rate slower than $r^{-3}$ is an approximation to the
true system, whose degree of accuracy must be quantified.  In this
view, a profile with an infinite total mass takes second seat to its
truncated version, regardless of the truncation mechanism.  This may
sound contrary to prevailing practice, but is absolutely necessary
to our goal because, in the theoretical framework, non-physical
profiles result in inaccurate lightrays.

Taking a given truncated profile as the ``true'' physical system,
there are thus two different approximations to quantify in relation
to the ``true'' mass of the system as predicted by the theoretical
approach:  the error introduced by the thin-lens approximation and
the further error introduced by the use of a profile that is
integrable along the line of sight but which results in an infinite
total mass.  In this paper we refer to the latter as the error
introduced by ``removal of truncation'', for lack of a better
phrase. Both errors are quantified in the present work within a
given family of density profiles.

One can, and probably should, think of the truncation mechanism
itself as an approximation: soft versus hard, for instance, or
different kinds of soft truncation, all applied to the same model.
In that case, it makes sense to quantify the error introduced by the
truncation mechanism as a third source of inaccuracy in the mass
estimates.  We are able to provide an estimate of the truncation
mechanism error by comparing mass predictions using soft and hard
truncation schemes for the same class of density profiles.

The gravitational lens scenario considered is the case of the
inference of lens mass from the observation of Einstein rings
(strong lensing).  The family of density profiles of interest is
described in Section~\ref{models:sec}. The mass prediction by the
conventional approach is discussed in Section~\ref{thin-lens:sec}.
The calculation of the ``true'' mass as predicted by the theoretical
framework is found in Section~\ref{geodesic:sec}. The results of our
study are found in Section~\ref{compare:sec}
Section~\ref{truncationmechanisms:sec}. We conclude in
Section~\ref{discussion:sec} with some remarks of interest that
follow from our study.


\section{Dark matter haloes} \label{models:sec}

Numerical modeling of dark matter haloes predicts a density profile
of the form \cite{NFW}
\begin{equation} \frac{\rho(r)}{\rho_{\mbox{\scriptsize crit}}}= \frac{\delta_c}
{\left(\frac{r}{r_s} \right) \left( 1 + \frac{r}{r_s} \right)^2},
 \label{NFW:rho}\end{equation}

\noindent hereafter referred to as the NFW model, where
$\rho_{\mbox{\scriptsize crit}}$ is the critical density (given by $
3H(z)^2/8\pi G$ in terms of the gravitational constant and Hubble's
constant at the lens redshift), $r_s$ is a scale radius defined as
the peak of $r^2\rho(r)$, and $\delta_c$ is a characteristic density
contrast. A
``virial'' radius is defined as the radius $r_{200}$ of the sphere
with mean density equal to $200\rho_{\mbox{\scriptsize crit}}$, and
is used to identify collapsed particle systems in numerical
simulations. The ratio $r_{200}/r_s\equiv c$ is referred to as the
halo's concentration. By definition of the virial radius one has, as
a consequence,

\begin{equation} \delta_c = \frac{200}{3}\frac{c^3}{\log(1+c)
- c/(1+c)}. \end{equation}

Falling off as $r^{-3}$, the NFW profile is not integrable over all
space, which is interpreted as an indication that the model fails at
large radius. Hence, a truncation mechanism is needed for long-range
applications, such as gravitational lensing. Two truncation
mechanisms for the NFW model have been used: hard truncation
\cite{takada,kf} and smooth truncation \cite{baltz}. The
hard-truncation mechanism consists of discontinuously terminating
the model by assuming that the mass density is identically vanishing
outside a given radius. In the lack of good knowledge pertaining to
halo edges, the choice of termination radius is largely arbitrary.
The common choice has been to terminate the NFW model at the virial
radius.

The smooth truncation mechanism of Baltz et al. \shortcite{baltz}
consists of multiplying the NFW profile by a factor that falls off
at least like $r^{-2}$, being close to 1 within small values of $r$.
This mechanism yields a sufficiently fast decay to ensure a finite
total mass (at least $r^{-5}$), while preserving the original
profile within a finite radius, within some accuracy.  The smoothly
truncated profile of Baltz et al. \shortcite{baltz} is

\begin{equation} \rho_n(r) = \frac{\delta_c \rho_{\mbox{\scriptsize crit}}}
{\left(\frac{r}{r_s} \right) \left( 1 + \frac{r}{r_s} \right)^2
\left(1 + \left(\frac{r}{r_t} \right)^2\right)^n}
\label{baltz:rho}\end{equation}

\noindent where $n$ is a positive integer and $r_t$ is a parameter
referred to as the tidal radius. For the purposes of the current
investigation, we restrict attention to the case $n=1$, hereafter
referred to as the truncated profile. For notational convenience we
use $\rho_0$ for the NFW model, since clearly the expression
(\ref{baltz:rho}) reduces to the NFW model (\ref{NFW:rho}) for
$n=0$. If we use a rescaled coordinate $x\equiv r/r_s$, the virial
radius would be located at $x=c$ and the tidal radius would be at
$x=\tau\equiv r_t/r_s$.  At any given radius $x$, the truncated
profile $\rho_1$ differs from the NFW profile $\rho_0$ by
\begin{equation}
\frac{\rho_1(x)-\rho_0(x)}{\rho_0(x)}=-\frac{x^2}{x^2+ \tau^2} .
\end{equation}

\noindent As a consequence, one way to interpret the tidal radius is
as the radius at which the profiles differ by 50\%.  At the virial
radius the profiles differ by
\begin{equation}\label{discrepancy}
\frac{\rho_1(c)-\rho_0(c)}{\rho_0(c)}=-\frac{1}{1+
(\frac{\tau}{c})^2}.
\end{equation}

\noindent Large values of the tidal radius $\tau$ relative to the
virial radius thus ensure that the truncated profile remains
sufficiently close to the NFW profile within the virial radius. For
the profiles to differ by 10\% or less at any point within the
virial radius, the tidal radius needs to be chosen as $\tau=3c$ or
larger. For this range of values of $\tau$, the virial mass (the
mass contained within the virial radius) of the truncated profile
differs from the virial mass of the NFW profile by less than 3\%.

On the other hand, for large values of the tidal radius the
truncated profile approaches the NFW profile, acquiring the
undesirable features of the NFW model, such as a very large
unrealistic total mass.  The optimal value of $\tau$ would thus be
the smallest possible value that yields the desired accuracy within
the virial radius, which is somewhat arbitrary itself.  Baltz et al.
\shortcite{baltz} justify the choice of $\tau=2c$ on the basis that
for this tidal radius the virial mass of the truncated profile
differs from the NFW virial mass by only 6\%. With this choice,
however, the profiles themselves differ by up to 20\% at the virial
radius, a difference that may be considered significant. At any
rate, the tidal radius should not be viewed as an independent
parameter but should be considered a function of the concentration
parameter.  With this assumption, the models have two free
parameters: the concentration parameter $c$ and the scale radius
$r_s$.

\section{Mass estimates by thin-lens gravitational lensing}\label{thin-lens:sec}

We are interested in estimating the error in the prediction of the
mass of a cluster by the observation of Einstein rings produced by
the perfect alignment of a source of light behind a massive
spherically symmetric deflector along the line of sight. In the
usual manner \cite{ehlers}, the radius $s$ of the Einstein ring on
the deflector's plane (perpendicular to the line of sight) at a
distance $D_l$ from us is the root of the lens equation:

\begin{equation}\label{TLequation} 1 -  \frac{M_n(s)}
{\Sigma_{\mbox{\scriptsize cr}}\pi s^2} = 0 .
\end{equation}

\noindent where $M_n(s)$ is the projected mass within a distance $s$
to the center of the lens:

\begin{equation}
 M_n(s) \equiv 2\pi \int_0^s s'ds'\int_{-\infty}^{\infty} \rho_n\left(\sqrt{s'^2+\ell^2}\right)
 d\ell .
\end{equation}

\noindent The constant $\Sigma_{\mbox{\scriptsize cr}}$ is the
characteristic surface mass density, given by
\begin{equation}
\Sigma_{\mbox{\scriptsize cr}}=\frac{c^2}{4\pi
G}\frac{D_s}{D_lD_{ls}},
\end{equation}

\noindent where $D_s$ is the distance to the source, $D_{ls}$ is the
distance between the lens and the source,  $G$ is the universal
gravitational constant, and $c$ is the speed of light, not to be
confused with the NFW concentration parameter.

For a given configuration of source and lens, different spherically
symmetric models $\rho_n$ will predict different sizes of Einstein
rings by virtue of the difference in their projected masses
$M_n(s)$. In principle, given a known Einstein ring and known source
and lens locations, equation (\ref{TLequation}) determines the
concentration $c_0$ or $c_1$ of the NFW model or the truncated
model, respectively, as functions of the scale radius $r_s$.
Different concentrations $c_0$ and $c_1$ would lead to different
predicted virial masses.  In practice, we fix the value of the scale
radius, the Einstein ring's angle and the distances to the lens and
the source, and determine the concentration by numerically solving
equation (\ref{TLequation}) for $c_n$ to high accuracy using a
standard bisection algorithm.

The value of the concentration parameter determines the virial mass
of the model $M_n^{\mbox{\scriptsize 200}}$ by:

\begin{equation}\label{virialmass}
M_n^{\mbox{\scriptsize 200}} \equiv 4\pi\int_0^{c_nr_s}\!\!r^2
\rho_n(r) dr
\end{equation}

\noindent which, in the case of $n=0$, is equivalent to $
M_0^{\mbox{\scriptsize 200}} \equiv c_0^3 \pi \frac{800}{3}
\rho_{\mbox{\scriptsize crit}} r_s^3 $.

\section{Mass estimates by numerical integration of the lightrays}\label{geodesic:sec}

The standard gravitational lensing formalism, leading to
Eq.(\ref{TLequation}), assumes that the lightrays from the source to
the observer are straight lines with sharp bending at the plane
containing the deflector. The bending and all its consequences are
obtained by ``projecting'' the mass density along the line of sight,
namely, by integrating the mass density along the line of sight.
This formalism is shadowed by a general relativistic formalism in
which the lightrays from the source to the observer curve according
to the geodesic equation of the metric of spacetime. The spacetime
in this case is a lens with a weak gravitational field located on a
expanding cosmological background. The lens is described by the
truncated model $\rho_1$.

There is some leeway as to how to put together the metrics of an
isolated lens and that of a Friedman-Robertson-Walker cosmology, in the
sense that, to our knowledge, a rigorous and complete derivation of the
Einstein equations valid for the particular case of an isolated lens on
a cosmological background is still lacking.  The Einstein equations
matter because the motion of the spacetime affects the matter density,
which affects the metric in turn.  So long as the rigorous results
remain outstanding, plausible (if unproven) descriptions of the
behavior of the matter in the lens may need to be used. A widely
accepted notion, supported by preliminary results, is that the matter
distribution in a \textbf{cluster of galaxies} does not partake of the
general expansion (namely, \textbf{clusters} do not expand internally).
This can be modeled with a metric that approaches the standard
cosmology at large distances from the lens, but which, in the vicinity
of the lens itself, is indistinguishable from an isolated lens in flat
space. Such a metric would be of the form

\begin{equation} ds^2 = (1+2 \varphi(r,t))dt^2
- a(t)^2 (1-2\varphi(r,t)) \{ dr^2 + r^2 (d\theta^2 + \sin^2\theta
d\phi^2 )\} . \label{m1} \end{equation}

\noindent where $\varphi$ represents the non-expanding isolated lens
centered at $r=0$.  Wherever the function $\varphi$ is not
negligibly small, the space part of the metric would be equivalent
to that of an isolated lens on flat space if $\varphi(r,t)$ is the
Newtonian potential of the lens in proper coordinates, namely
$\varphi(r,t) = U(a(t)r)$ with $\nabla^2U = 4\pi G \rho_1$.  By
requiring the solution of this equation to vanish at infinity and be
finite at the origin, the Newtonian potential $U$ of the mass
distribution $\rho_1$ is found to be

\begin{equation}
U(x) = \frac{Gm}{r_s}\frac{\tau^2}{(1+\tau^2)^2} \Bigg[
 \arctan (x/\tau) \Big( \frac{1}{\tau}-\tau-2\frac{\tau}{x} \Big)
+\ln\left(\frac{1+(x/\tau)^2}{(1+x)^2}\right)\Big(
\frac{\tau^2-1}{2x} -1\Big) +\frac{\pi(\tau^2-1)}{2\tau}
-2\ln\tau\Bigg] \label{baltz:phi1}
\end{equation}

\noindent with $x\equiv r/r_s$. Here  $m \equiv
4 \pi r_s^3 \delta_c\rho_{\mbox{\scriptsize crit}}$.

The function $\varphi(r,t)$ is obtained by the substitution
$r\rightarrow a(t) r$ in Eq. (\ref{baltz:phi1}), and is, in principle,
a time-varying function.   However, the time variation is extremely
slow in the conditions of observable lensing, where the source and
observer are very far away from the lens. \textbf{Since $\varphi$ is
relevant only in the vicinity of the lens, one could substitute the
time-varying $a(t)$ with a constant $a(t_l)$ with $t_l$ being  a
representative time at which the lightray being considered reaches the
vicinity of the lens.} A posteriori, we find that the results of the
numerical calculations that follow are not significantly affected by
the choice of fixing the value of $a(t)$ to a constant, so we follow
common practice and set

\begin{equation}
\varphi(r,t) = U(a(t_l)r)
\end{equation}

\noindent for use in the metric (\ref{m1}). This has the additional
advantage that in these conditions the metric (\ref{m1}) is
conformally static, as assumed in Schneider et al
\shortcite{ehlers}, and thus the well known phenomena associated
with lenses follow as per that text,  including Eq. 6.

In the standard cosmology, the expansion parameter $a(t)$ is given
by

\begin{equation} a(t) = \left( \frac{\Omega_m}{\Omega_\Lambda}
\right)^{1/3} \left\{ \sinh\left( \frac{3 H_0 \sqrt{\Omega_\Lambda}
t}{2} \right) \right\}^{2/3}, \label{a1}  \end{equation}

\noindent with $H_0 = 70$~km/s/Mpc, $\Omega_m = 0.3$ and
$\Omega_\Lambda = 0.7 = 1- \Omega_m$.

Since the metric is spherically symmetric, we can restrict attention
to the equatorial plane without loss of generality. The null
geodesic equations are equivalent to the the Euler-Lagrange
equations of the Lagrangian

\begin{equation}  {\mathcal{L}} = (1+2 \varphi)
\dot t^2 - a^2(t) (1-2\varphi) \{ \dot r^2 + r^2 \dot \phi^2 \}
 = 0 , \label{lagrangian} \end{equation}

\noindent  The Lagrangian is set to zero because the geodesics are
null.  The Euler-Lagrange equations are equivalent to five
first-order ODEs, which we can write as

\begin{eqnarray}  \dot t &=& v_t \nonumber \\
\dot r & = & v_r \nonumber\\
\dot v_t & = & - \frac{da}{dt} \frac{v_t^2}{a}  - 2(1-2\varphi)
\frac{\partial
\varphi}{\partial r} v_t v_r \nonumber\\
\dot v_r &=& -\frac{ 2 v_t v_r}{a} \frac{da}{dt} + 4
\frac{\varphi}{a^2} \frac{\partial \varphi}{\partial r} v_t^2
\nonumber \\ && - \frac{2b^2}{a^4 r^2} \frac{\partial
\varphi}{\partial r} (1+ 6\varphi) + \frac{b^2}{a^4 r^3}(1+4\varphi)
\nonumber \\\dot \phi &=& -\frac{b}{a^2 r^2} (1+ 2\varphi).
\label{ODES} \end{eqnarray}

\noindent These equations are accurate to first order in $\varphi$.
The parameter $b$ is a constant of integration and is related to the
``observation angle'' at the observer, or the angle between the
lightray and the optical (radial) axis connecting the observer to
the lens as explained in Kling \& Frittelli \shortcite{kf}:

\begin{equation} \sin\theta_{obs} = \theta_{obs} =
\frac{b}{r_0}(1+2\varphi_0), \label{ring} \end{equation}

\noindent where the potential $\varphi_0$, is evaluated at the
observer position, $r_0$.

By integrating the equations of motion from a given point, for a
specified set of parameters $(r_s,c)$ describing the matter density,
one finds a specific light ray passing through that point.  Our
scheme for a mass prediction is based on lightrays making Einstein
rings, for which the source, lens and observer are collinear and are
at known distances from each other. So the beginning and end point
of the null geodesics being sought are known. In these conditions,
for every observed Einstein ring $\theta_{obs}$, as per
Eq.~\ref{ring}, there is an associated set of values of $(r_s,c)$
that are consistent with the observed angle as well as the fixed
configuration of lens, source and observer. As in the case of the
prediction by the thin-lens gravitational lens formalism, in
practice we fix $r_s$ and find $c$ using Newton's method and a ray
shooting technique.  The ray shooting technique consists essentially
of integrating rays for a given value of $c$ and seeing where they
land with respect to the end point aimed at, either in front of it
or behind it, subsequently adjusting $c$ accordingly, until the ray
integrated lands sufficiently close to the target, within a
specified accuracy.  In practice, the integration is done from
observer to source.

For the purposes of integrating the null geodesics, we re-scale the
time and radial coordinates by the age of the universe. The
equations of motion are integrated using an adaptive step-size
Runge-Kutta-Fehlberg 4-5 method based on the implementation in
\cite{nrc}.  The numerical error is at least 500 times smaller than
the difference between the result by null geodesic integration and
the prediction by the thin-lens formalism.

The value of $c$ found by this method is considered to be the
``true'' value of the model and is used to predict the ``true''
virial mass of the model $M^{\mbox{\scriptsize 200}}$ by Eq.
(\ref{virialmass}) with $n=1$, namely:

\begin{equation}
M^{\mbox{\scriptsize 200}} \equiv 4\pi\int_0^{c r_s}\!\!r^2
\rho_1(r) dr .
\end{equation}


\section{Comparison of the thin-lens mass prediction relative to
the null-geodesic prediction} \label{compare:sec}

For our first study we arbitrarily choose a series of five reasonable
lensing systems with varying redshifts, and calculate the concentration
parameters $(c, c_1, c_0)$ by the method of null-geodesics, thin-lens
with truncation and thin-lens with no truncation, respectively. We do
this for $\tau=2c$ and for $\tau=3c$ separately. In all cases, the
scale radius is arbitrarily assumed to be $r_s=0.25$ Mpc. The results
are shown in Tables \ref{t3c:table} and \ref{t2c:table}.

In all cases, the concentration parameter $c_1$ is slightly higher than
the true value $c$, whereas $c_0$ is closer to and generally lower than
$c$.  \textbf{However, the reader should keep in mind that when the
virial mass is used as the parameter for comparison, the relative
values of $c_0$ and $c_1$ are not significant in themselves.}  The
tables show for each case the errors in the virial masses of the
truncated and non-truncated thin lens models relative to the true model
(which is truncated with no assumption of thin lenses).

We can verify that, whether truncation is applied or not, the use of
the thin-lens approximation overestimates the virial mass of the
system, as the relative errors are negative in all cases.

The sizes of the two separate simplifying assumptions (thin lens and
removal of truncation) are revealed to be very different.  The error
incurred by the sole simplifying assumption of thin lenses,
displayed in the column headed by $1-M_1^{200}/M^{200}$, is
generally between 0.3\% and 0.7\% for $\tau=2c$, and between 0.2\%
and 0.4\% for $\tau=3c$.

\begin{figure}
\hbox{\hbox{\includegraphics[width=6in,angle=0]{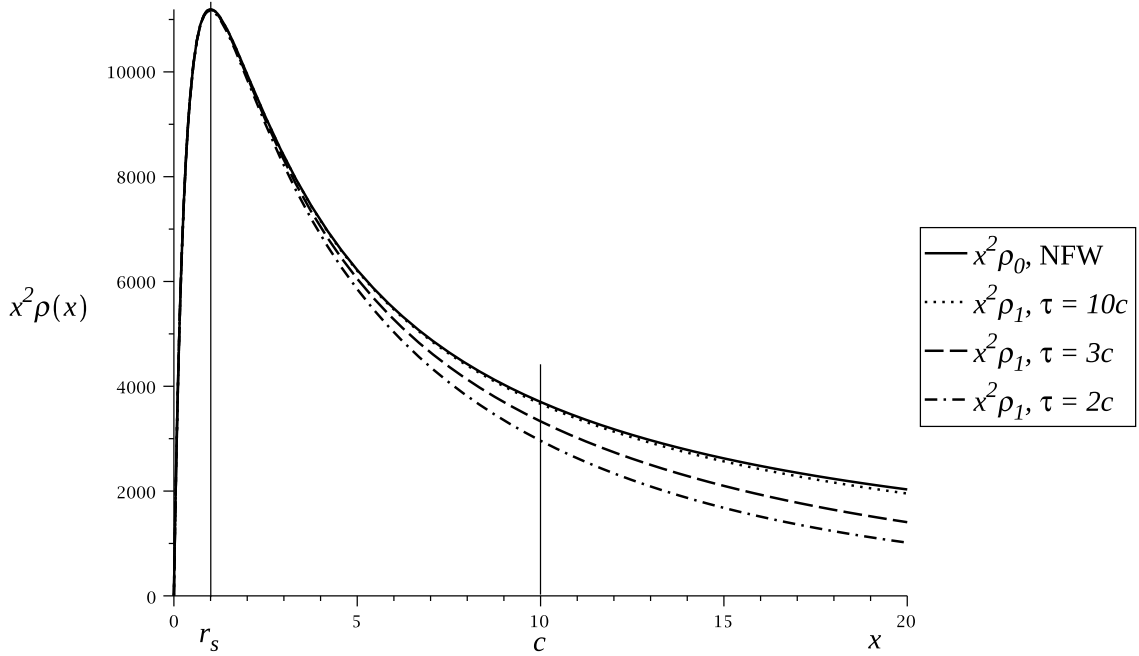}} }
\caption{\label{profiles:fig} Density profiles as functions of the
scaled distance $x\equiv r/r_s$. The solid line represents the
standard NFW profile $x^2\rho_0(x)$, with a non-integrable fall-off.
The other three lines represent the $n=1$ truncated profiles
$x^2\rho_1(x)$ for three different values of the tidal parameter
$\tau$ as multiples of the concentration $c$.  All three truncated
profiles have integrable fall-off rates. However, the larger the
ratio $\tau/c$, the closer the truncated profile matches the NFW
profile within the virial radius, represented by $c$ as a scaled
distance.}
\end{figure}

By contrast, the error incurred by further removing the truncation in
order to simplify the estimation of the mass is about 6\% for $\tau=2c$
and 3\% for $\tau=3c$. This is the dominant source of error in the
virial mass.  Keeping in mind that ``truncation'' in this model refers
merely to a fast decay, this is not surprising, as different decay
rates lead to significantly different profiles within the virial
radius, as shown in Figure \ref{profiles:fig}.

For our second study, we consider three observed systems with
symmetrical arcs where one can use an Einstein ring analysis to
determine parameters of the system with confidence. The systems are
RXJ1347-1145, the brightest source in the ROSAT all sky survey
\cite{RXJstrong}; MS2137-23, a relatively structure-free, four-arc
system \cite{ms2137}; and cluster A of the high-redshift cluster
RCS2319+00 \cite{rcs2319}. Table \ref{systems:table} gives the
redshift and arc distribution details of these clusters.

\begin{figure}
\hbox{\hbox{\includegraphics[width=3.5in,angle=0]{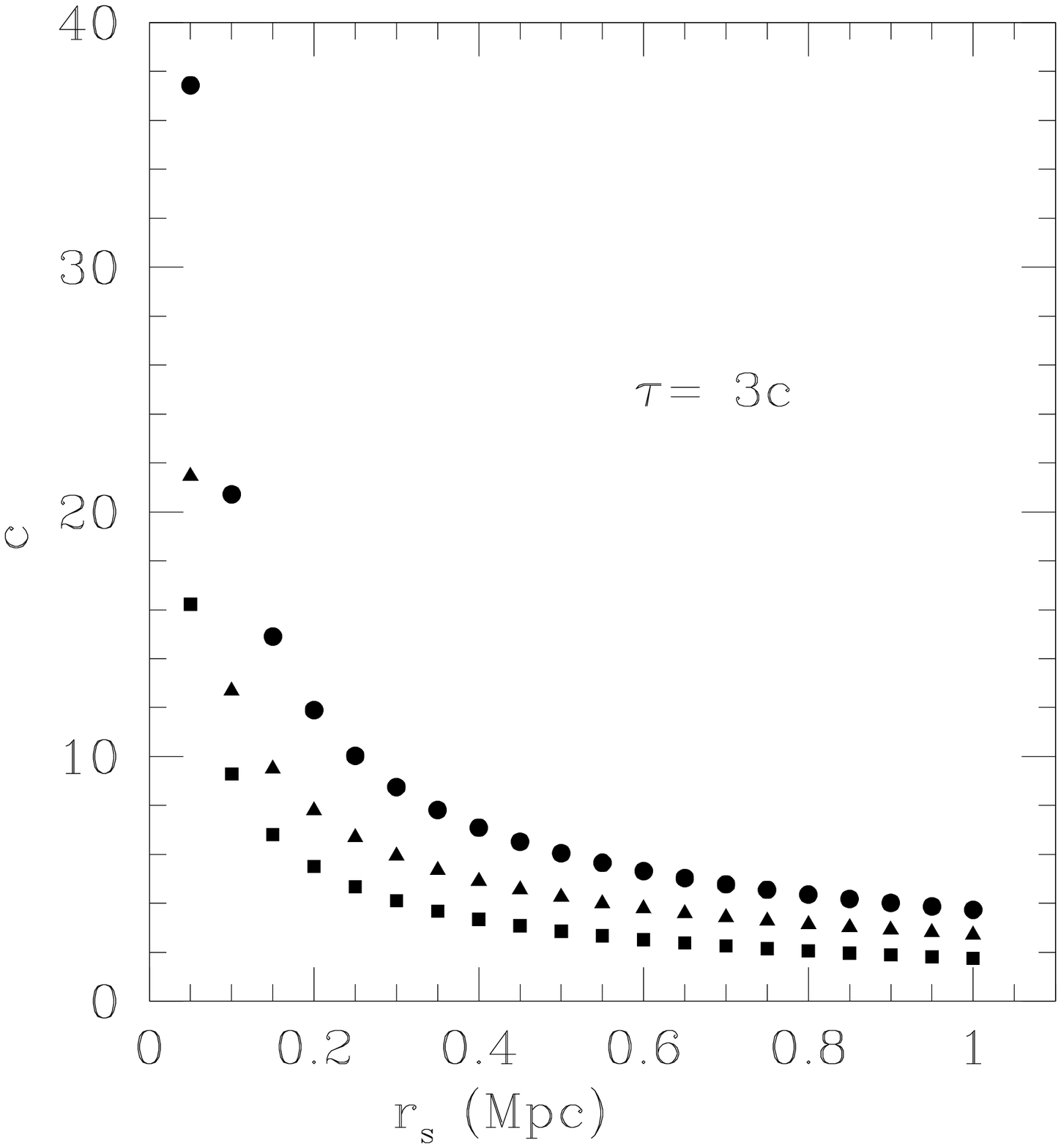}}
      \hbox{\includegraphics[width=3.5in,angle=0]{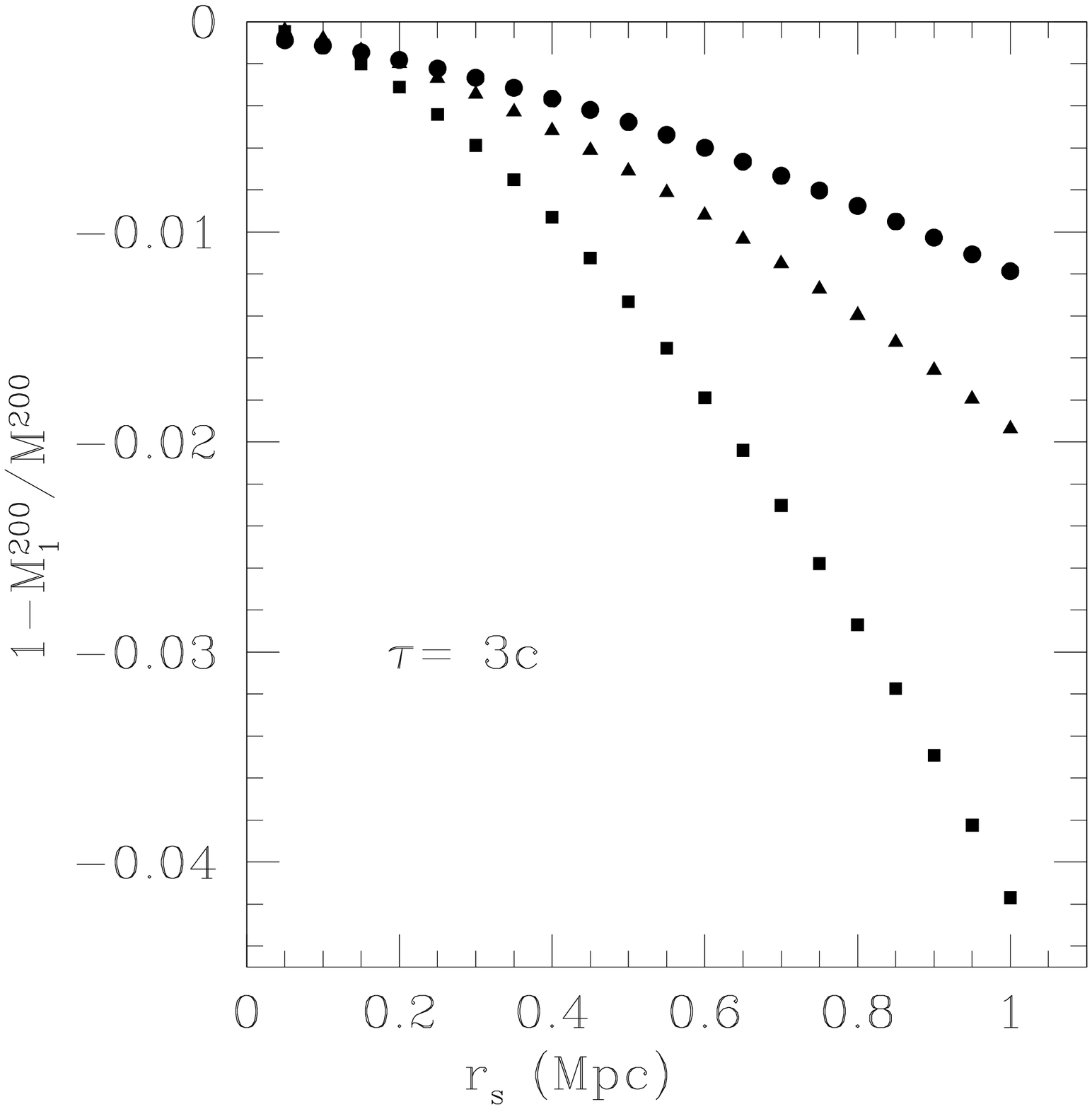}}
      }
\caption{\label{t3c:fig} Contour in $(r_s, c)$ space (left) and
relative error between thin lens prediction and actual total mass
(right) for the three lensing systems modeled as truncated
distributions $\rho_1$ with $\tau=3c$. Circles correspond to the 35
arc sec arcs in RXJ1347-1145, triangles correspond to the 16 arc sec
arcs in MS 2137-23, and squares correspond to the 12 arc sec arcs
around cluster A of RCS 2319+00.}
\end{figure}

For each one of these three systems, the Einstein rings are known.
Modeling each system as a truncated $\rho_1$ density profile with
$\tau=3c$ leads to two unknown parameters for each system: the value
of the concentration parameter $c$ and the value of the scale radius
$r_s$.  The left panel of Figure \ref{t3c:fig} shows the contour
plot in the $r_s - c$ plane for the Einstein rings of these three
systems. For this plot, $r_s$ was varied arbitrarily and $c$ is the
true concentration parameter determined using the null-geodesic
integration formalism. In the right panel of Fig. \ref{t3c:fig}, we
show the error in the virial mass $1-M_1^{200}/M^{200}$. As the
errors are negative, we verify that the thin-lens approximation
overestimates the mass of each system.  For reasonable values of
$r_s$ and $c$, the errors are of the order of $0.5\%$.

\begin{figure}
\hbox{\hbox{\includegraphics[width=3.5in,angle=0]{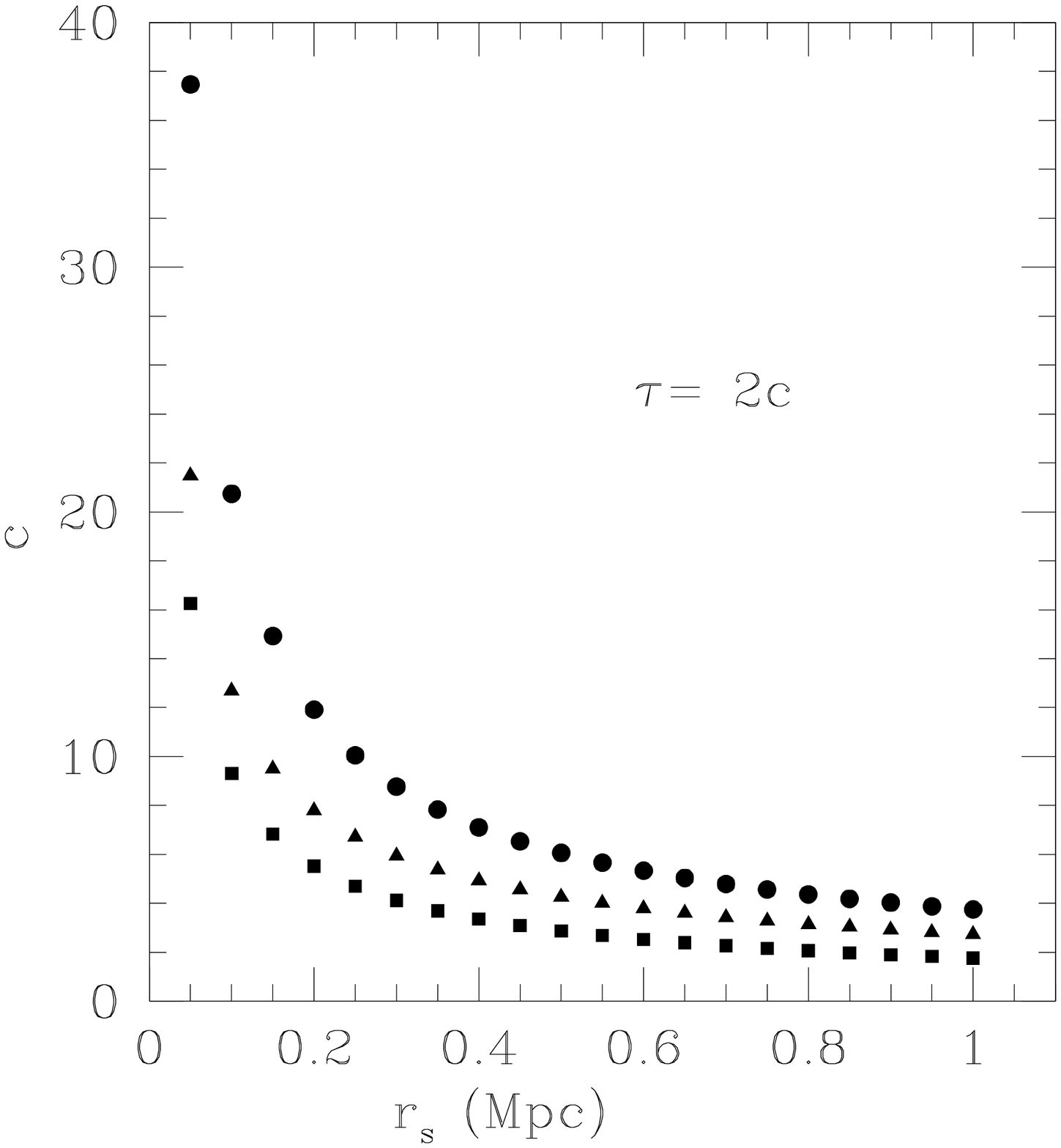}}
      \hbox{\includegraphics[width=3.5in,angle=0]{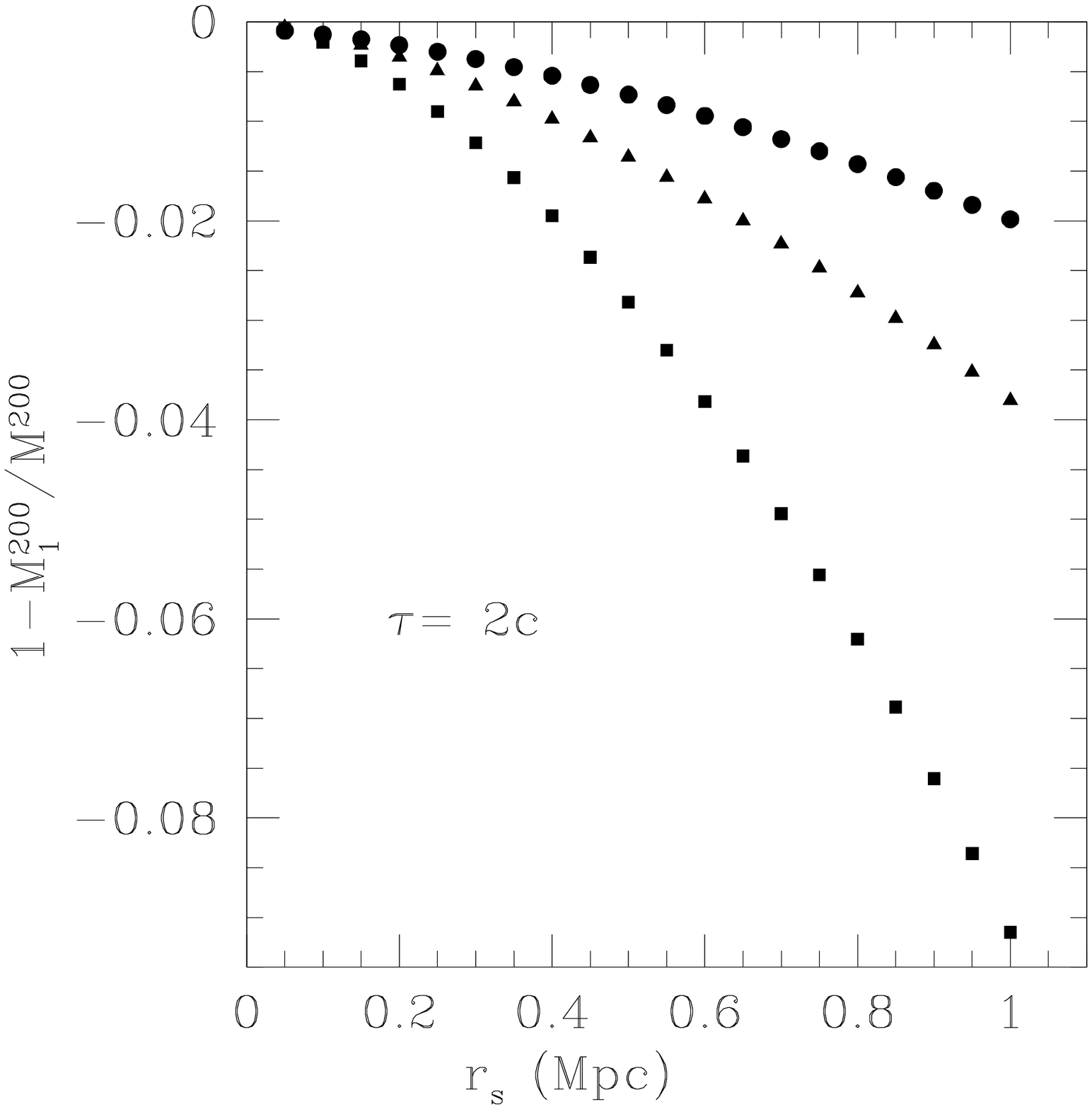}}
      }
\caption{\label{t2c:fig} Contour in $(r_s, c)$ space (left) and
relative error between thin lens prediction and actual total mass
(right) for three lensing systems modeled as as truncated
distributions $\rho_1$ with $\tau=2c$. Circles correspond to the 35
arc sec arcs in RXJ1347-1145, triangles correspond to the 16 arc sec
arcs in MS 2137-23, and squares correspond to the 12 arc sec arcs
around cluster A of RCS 2319+00.}
\end{figure}

Figure \ref{t2c:fig} shows the corresponding results for the case that
the modeling is done with $\tau=2c$.  One can verify that even though
the concentration parameters do not differ significantly from the ones
in the case of $\tau=3c$, the accuracy of the mass prediction is
reduced by a factor of 2, as the maximum relative error in the mass is of the
order of 10\%.


\section{Comparison of truncation mechanisms} \label{truncationmechanisms:sec}

As indicated in the introduction, one of the fundamental problems of
gravitational lensing is the level of uncertainty in the underlying
physical model of the lens irrespective of the approximation method
used to make predictions. Here we address some of the consequences
of this uncertainty by calculating the difference between the virial
masses predicted exactly (with null geodesics) for two different
models assumed to represent the same physical lens.

The two models differ in the truncation mechanism applied to an NFW
lens: soft versus hard truncation. We use a density profile $\rho_1$
with $\tau=3c$ as a softly truncated NFW lens with $r_s=0.25$ Mpc. The
hard truncation model is a density profile $\rho_0$ with $r_s=0.25$ Mpc
within the virial radius, and zero outside of the virial radius. The
hard truncation model is discussed extensively in our previous work
\cite{kf}, where the gravitational potential needed for the integration
of the null geodesics is found (Equation (34) in \cite{kf}).

Table \ref{comparison:table} shows the result of this study.  The
predicted virial masses for both models differ within $4\%$.  This is
an independent indication that the truncation mechanism itself is much
more important than the simplifying assumption of thin lenses, and is
the dominant source of error in mass estimates by gravitational
lensing, by an order of magnitude. Perhaps counter-intuitively, the
difference in profiles outside of the virial radius (as seen in Figure
\ref{profiles:fig}) is a significant source of inaccuracy in the
prediction of virial mass by actual light paths traveled.

One puzzling result is that the hard truncation model consistently
overestimates the mass with respect to the soft truncation model (as
indicated by the negative values in the last column in Table
\ref{comparison:table}). With both profiles being almost identical
within the virial radius, one would expect that the soft truncation
mechanism (just be virtue of having more total mass) would produce
larger bending and would lead to a higher estimate for the virial mass.


\section{Discussion} \label{discussion:sec}

One of the main results obtained through our present study is that
given a reasonable physical mass profile, such as $\rho_1$, the use of
the thin-lens approximation as a simplifying assumption to predict the
virial mass of the system leads to very small errors, which we found to
be of order 0.5\%.   This can be interpreted as a validation of the use
of the thin-lens approximation in the present context and, by
extension, as an indication of the validity of the thin-lens
approximation in generic physically reasonable systems.

A second result found is that\textbf{, for this family of models,}
given a reasonable physical mass profile, such as $\rho_1$, removing
the truncation as a further simplifying assumption (such as using the
NFW) in addition to the thin-lens approximation leads to a significant
loss of accuracy in the prediction of the virial mass of the system in
the cases $\tau=2c$ and $\tau=3c$.  The error introduced by removing
the truncation is larger by a factor of at least 10 over the error
introduced by the use of the thin-lens approximation.

For practitioners of the thin-lens approximation, one of the main
concerns with the soft truncation mechanism studied here is the
built-in $\tau$-dependent discrepancy in the mass profile.  In effect,
systems modeled with profiles $\rho_1$ with different values of
$\tau/c$ can vary significantly.  Because the truncation is effected by
means of a modification of the decay rate, the profiles diverge over a
significant range of distances.  What is important, however, is that
the $\tau-$profiles are reasonably close to the NFW profile $\rho_0$
within the virial radius.  This is because the NFW profile is known to
be accurate up to the virial radius, so any loss of accuracy within the
virial radius is a disadvantage of the $\tau-$dependent model.  Baltz
et al \shortcite{baltz} propose that a discrepancy of about 20\% in the
mass profile between $\rho_1$ and $\rho_0$ is acceptable, as the
penalty for working with a small tidal parameter $\tau=2c$.

However, larger values of the tidal parameter lead to better accuracy
in the mass profile within the virial radius. We have argued and
demonstrated that a slightly higher value $\tau=3c$ cuts the error in
both the virial mass and density profile by half. Raising the value of
$\tau$ relative to $c$ increases the accuracy of the profile, in
principle arbitrarily. However, a profile $\rho_1$ for high $\tau$ can
approach $\rho_0$ too accurately to be useful, if the predicted virial
mass $M_1^{200}$ approaches $M_0^{200}$ more accurately than the
limiting accuracy of the thin-lens approximation itself.  In other
words, since the thin-lens approximation itself carries a base accuracy
of 0.5\% in the virial mass prediction, there is no use in fine-tuning
the density profile to lead to an accuracy better than 0.5\% in the
virial mass.

Our present study thus allows us to formulate a criterion for an
optimal value of the tidal parameter $\tau$ relative to $c$:  $\tau/c$
should be as large as possible but larger than required to achieve an
accuracy of about $0.5\%$ in the virial mass $M_1^{200}$ relative to
the NFW virial mass $M_0^{200}$.  We can come up with a rough
estimation by noticing that 20\% accuracy in the profile leads to 6\%
accuracy in the virial mass, whereas 10\% accuracy in the profile leads
to 3\% accuracy in the virial mass.  The gross trend would indicate
that 0.1\% accuracy in the profile would achieve roughly 0.3\% accuracy
in the virial mass.  By equation \ref{discrepancy}, a relative error of
no less than 1\% in the profile leads to a tidal radius no larger than
$\tau\sim 10 c$. A value of the order of $\tau=10 c$ thus would
guarantee that the NFW profile is modeled with $\rho_1$ within the
virial radius to the limit of accuracy of the thin lens approximation,
while ensuring that the model has a finite total mass.

This rough estimate is validated by Table \ref{t10c:table} where the
errors in the virial masses $M_1^{200}$ and $M_0^{200}$ relative to the
true virial mass $M^{200}$ are shown in the last two columns.  The
column headed by $1-M_1^{200}/M^{200}$ represents the error incurred by
the use of the thin lens approximation, and is generally of order of
0.2\%.  The last column, headed by $1-M_0^{200}/M^{200}$, represents
the combined error of the thin lens approximation and lack of
truncation, and is generally of order 0.4\%.  The difference between
the values in both columns is the error introduced by the truncation
mechanism alone, and is of order 0.2\%, namely, comparable to the
thin-lens approximation error, as anticipated.  We claim that for
thin-lens practitioners, using $\tau$ much greater than $10c$ would be
inefficient, as it would lead to a truncation error smaller than the
error inherently built into the thin-lens approximation itself.

\begin{figure}
\hbox{\hbox{\includegraphics[width=2.3in,angle=0]{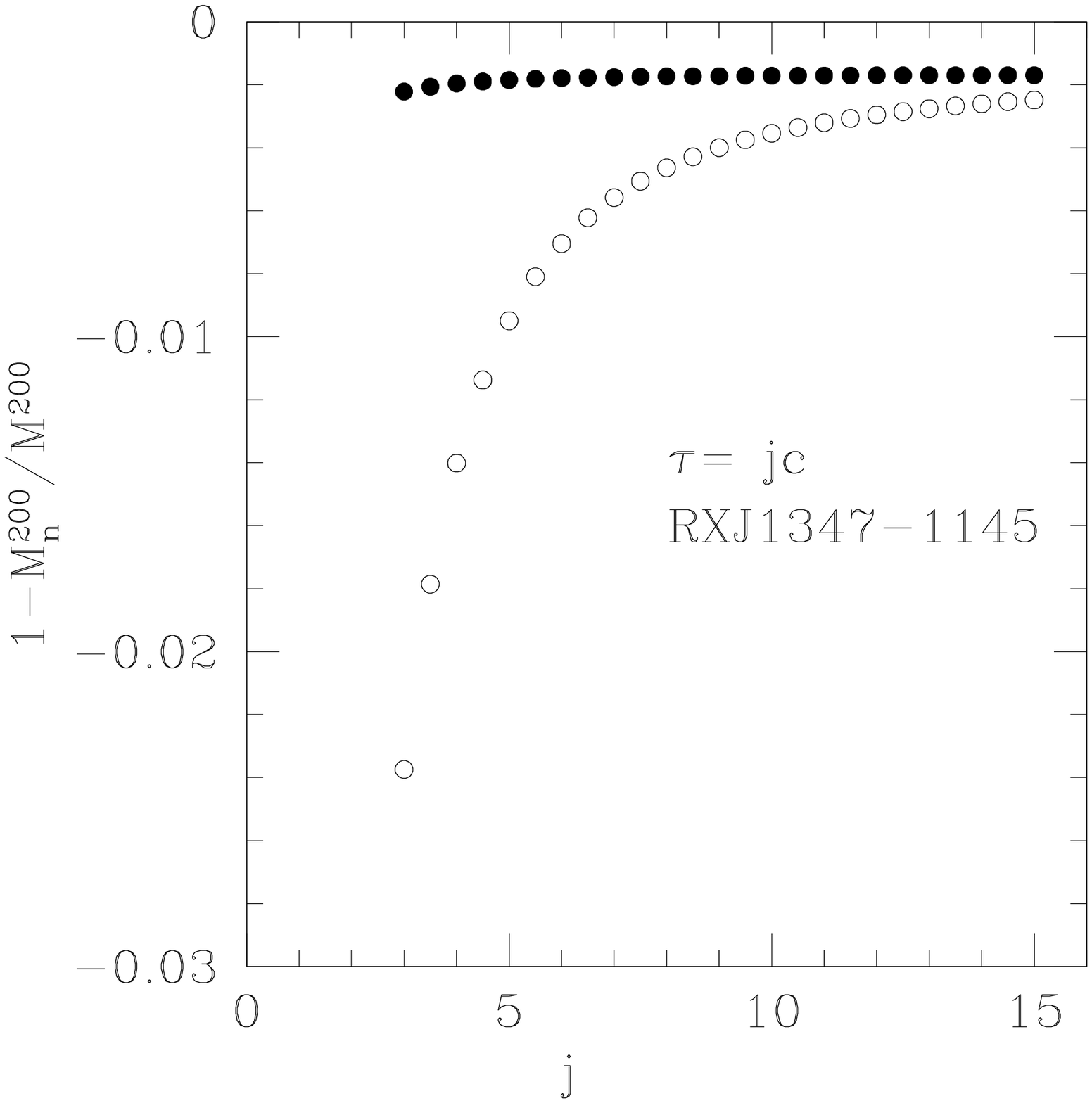}}
      \hbox{\includegraphics[width=2.3in,angle=0]{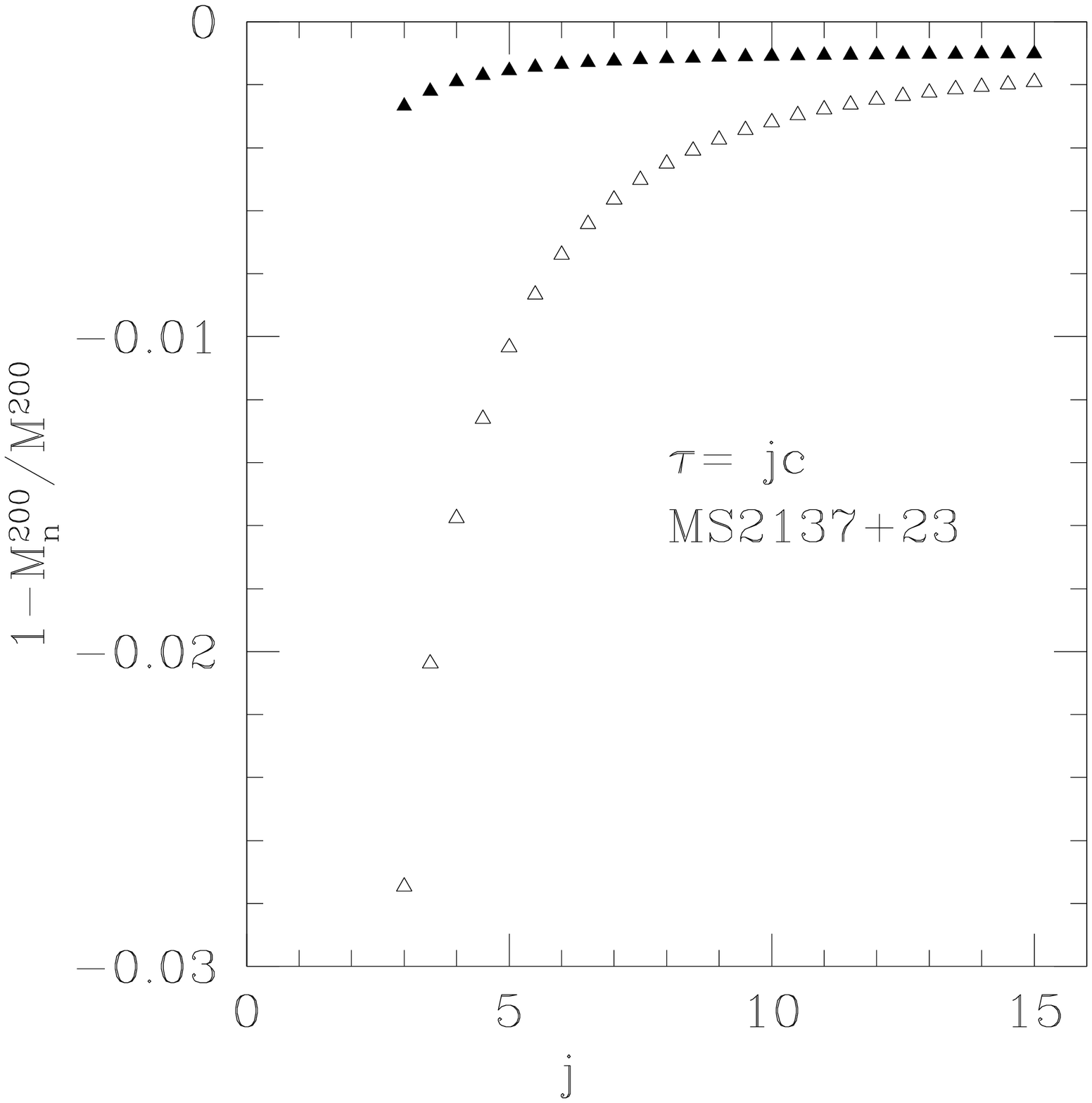}}
      \hbox{\includegraphics[width=2.3in,angle=0]{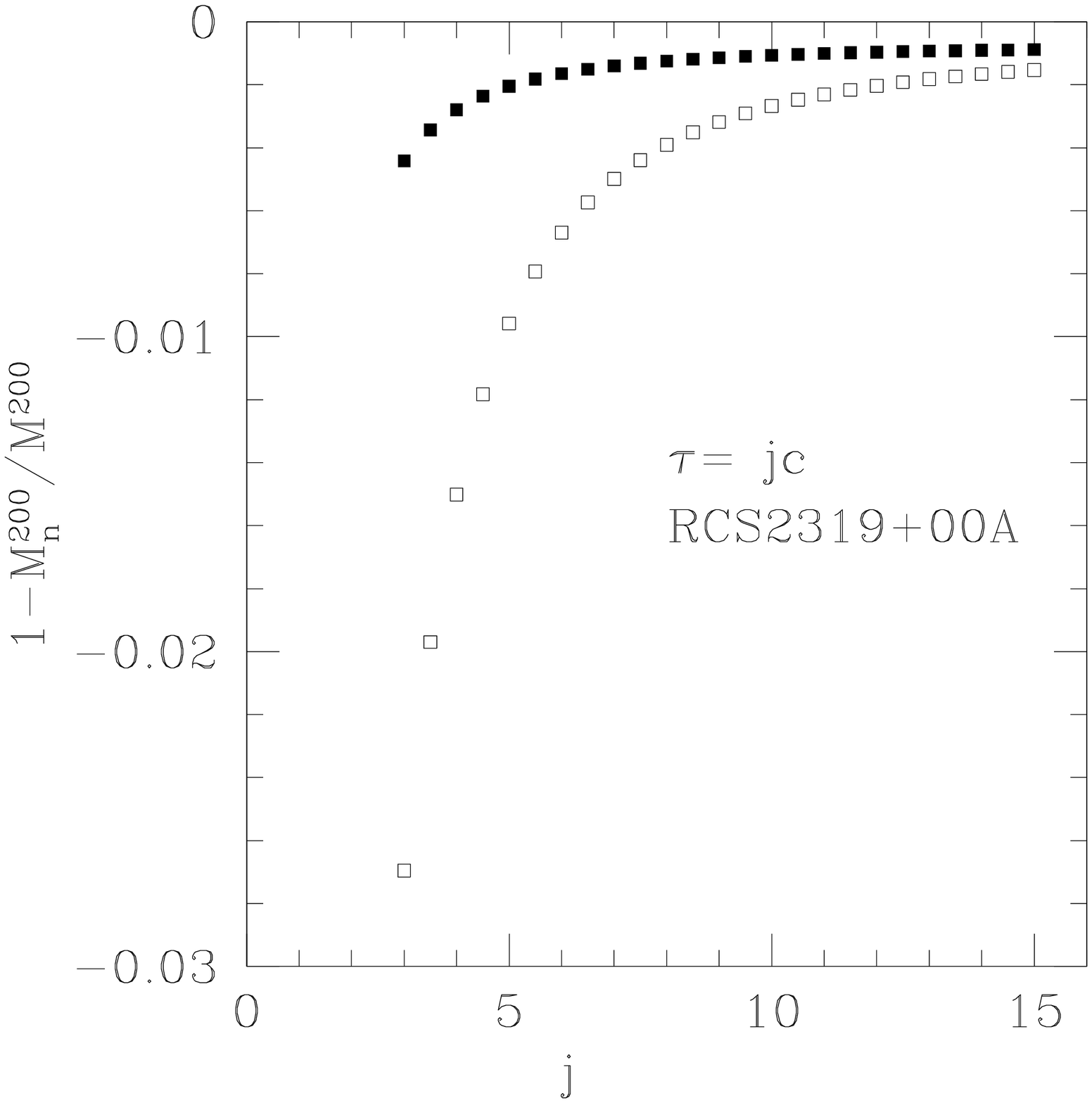}}
      }
\caption{\label{j:fig} Relative error between thin lens prediction
and actual total mass for three lensing systems for increasing
values of $\tau$ as multiples of the concentration parameter ($\tau$
= jc). Filled symbols correspond to the truncated model $\rho_1$ and
open symbols correspond to the non-truncated NFW model $\rho_0$.}

\end{figure}

Further validation of this claim is provided by Figure \ref{j:fig},
where the relative virial mass errors $1-M_1^{200}/M^{200}$ and
$1-M_0^{200}/M^{200}$ are calculated for values of $\tau=jc$ with
$j=1, 2, ... 15$, for the three observed systems of Table
\ref{systems:table}.  One can see that in all three cases the
thin-lens errors hover around 0.2\%, whereas the combined
truncation/thin-lens error drops dramatically from roughly 2\% at
$\tau=3c$ to less than 0.5\% at $\tau=8c$, reaching within a factor
of two of the thin-lens error at $\tau=10c$, indicating that the
truncation error alone is of the same size as the thin-lens error at
$\tau=10c$, as anticipated.

Other features to note from inspection of Tables \ref{t3c:table},
\ref{t2c:table} and \ref{t10c:table} are as follows.  The larger the
predicted mass, the more accurate the thin-lens prediction, with or
without truncation.  Nevertheless, the non-truncated prediction
improves only marginally in accuracy, in contrast to the truncated
prediction, which improves significantly.

\section{acknowledgements}

This material is based upon work supported by the National Science
Foundation under Grant Nos. PHY-0244752 and PHY-0555218. We
gratefully acknowledge the hospitality and support of the American
Institute of Mathematics during the progress of the workshop
``Gravitational Lensing in the Kerr Geometry,'' AIM, Palo Alto, July
5-10, 2005.

%
%

\clearpage

\begin{table} \begin{center}\begin{tabular}{rrcccccrr}
\hline $\tau=3c$ & & & & & & & \\
\hline System &  $\theta_E$ & $r_{200}$ (Mpc) &
$c$ & $c_1$ & $c_0$ & $M^{\mbox{\scriptsize 200}}~(10^{15} M_\odot)$ & $1-M^{\mbox{\scriptsize 200}}_1/M^{\mbox{\scriptsize 200}}$ & $1-M^{\mbox{\scriptsize 200}}_0/M^{\mbox{\scriptsize 200}}$ \\
\hline ~&~&~&~&~&~&~&~&~\\
~           & 10.0 & 2.05 &  8.201 &  8.210 &  8.204 & 1.14 & -0.0033 & -0.0289 \\
$z_l = 0.2$ & 17.5 & 2.29 &  9.153 &  9.161 &  9.154 & 1.60 & -0.0028 & -0.0274 \\
~           & 25.0 & 2.48 &  9.939 &  9.948 &  9.940 & 2.04 & -0.0026 & -0.0264 \\
$z_s = 0.4$ & 32.5 & 2.66 & 10.639 & 10.648 & 10.639 & 2.50 & -0.0025 & -0.0257 \\
~           & 40.0 & 2.82 & 11.282 & 11.291 & 11.281 & 2.98 & -0.0025 & -0.0251 \\
\hline
~           & 10.0 & 1.69 &  6.756 &  6.763 &  6.755 & 0.72 & -0.0031 & -0.0289 \\
$z_l = 0.3$ & 17.5 & 1.91 &  7.658 &  7.664 &  7.655 & 1.04 & -0.0025 & -0.0270 \\
~           & 25.0 & 2.10 &  8.415 &  8.421 &  8.410 & 1.38 & -0.0022 & -0.0257 \\
$z_s = 0.8$ & 32.5 & 2.27 &  9.093 &  9.100 &  9.087 & 1.74 & -0.0021 & -0.0248 \\
~           & 40.0 & 2.43 &  9.720 &  9.727 &  9.712 & 2.12 & -0.0020 & -0.0240 \\
\hline
~           &  5.0 & 1.34 &  5.376 &  5.384 &  5.377 & 0.36 & -0.0045 & -0.0319 \\
$z_l = 0.3$ &  9.0 & 1.49 &  5.972 &  5.979 &  5.971 & 0.50 & -0.0035 & -0.0297 \\
~           & 13.0 & 1.61 &  6.453 &  6.459 &  6.450 & 0.62 & -0.0029 & -0.0283 \\
$z_s = 1.5$ & 17.0 & 1.72 &  6.874 &  6.880 &  6.870 & 0.76 & -0.0026 & -0.0273 \\
~           & 21.0 & 1.81 &  7.258 &  7.264 &  7.253 & 0.88 & -0.0024 & -0.0265 \\
\hline
~           & 10.0 & 2.05 &  8.198 &  8.207 &  8.199 & 1.82 & -0.0032 & -0.0280 \\
$z_l = 0.6$ & 17.5 & 2.37 &  9.466 &  9.475 &  9.464 & 2.80 & -0.0029 & -0.0262 \\
~           & 25.0 & 2.64 & 10.546 & 10.556 & 10.544 & 3.86 & -0.0029 & -0.0251 \\
$z_s = 0.8$ & 32.5 & 2.88 & 11.521 & 11.533 & 11.518 & 5.04 & -0.0030 & -0.0243 \\
~           & 40.0 & 3.11 & 12.426 & 12.439 & 12.422 & 6.32 & -0.0032 & -0.0237 \\
\hline
~           &  5.0 & 1.30 &  5.205 &  5.214 &  5.205 & 0.74 & -0.0049 & -0.0313 \\
$z_l = 1.0$ &  9.0 & 1.48 &  5.933 &  5.941 &  5.930 & 1.10 & -0.0038 & -0.0286 \\
~           & 13.0 & 1.63 &  6.538 &  6.546 &  6.533 & 1.48 & -0.0033 & -0.0269 \\
$z_s = 1.5$ & 17.0 & 1.77 &  7.078 &  7.086 &  7.071 & 1.88 & -0.0030 & -0.0257 \\
~           & 21.0 & 1.89 &  7.576 &  7.584 &  7.567 & 2.30 & -0.0029 & -0.0247 \\
\hline
\end{tabular} \caption{The concentration parameters and relative
errors in the virial mass for a set of hypothetical lensing systems
assuming that the true system has a density profile $\rho_1$ with a
scale radius of $r_s=0.25$ Mpc and $\tau=3c$. We give the true $c$
value, and values obtained by thin lens approximations using a
density profile $\rho_1$  ($c_1$) and an NFW profile $\rho_0$
($c_0$). The virial radius quoted, $r_{\mbox{200}}$, is the one
calculated using the true $c$ value. The true mass,
$M^{\mbox{\scriptsize 200}}$ is given in units of $10^{15}$ solar
masses. \label{t3c:table} }
\end{center}
\end{table}
\clearpage

\begin{table} \begin{center}\begin{tabular}{rrcccccrr}
\hline $\tau=2c$ & & & & & & & \\
\hline System &  $\theta_E$ & $r_{200}$ (Mpc) &
$c$ & $c_1$ & $c_0$ & $M^{\mbox{\scriptsize 200}}~(10^{15} M_\odot)$ & $1-M^{\mbox{\scriptsize 200}}_1/M^{\mbox{\scriptsize 200}}$ & $1-M^{\mbox{\scriptsize 200}}_0/M^{\mbox{\scriptsize 200}}$ \\
\hline ~&~&~&~&~&~&~&~&~\\
~           & 10.0 & 2.05 &  8.202 &  8.216 &  8.204 & 1.12 & -0.0049 & -0.0611 \\
$z_l = 0.2$ & 17.5 & 2.29 &  9.156 &  9.168 &  9.154 & 1.54 & -0.0040 & -0.0579 \\
~           & 25.0 & 2.49 &  9.944 &  9.956 &  9.940 & 1.98 & -0.0035 & -0.0558 \\
$z_s = 0.4$ & 32.5 & 2.66 & 10.647 & 10.656 & 10.639 & 2.44 & -0.0033 & -0.0541 \\
~           & 40.0 & 2.82 & 11.289 & 11.300 & 11.281 & 2.90 & -0.0031 & -0.0528 \\
\hline
~           & 10.0 & 1.69 &  6.757 &  6.770 &  6.755 & 0.70 & -0.0054 & -0.0626 \\
$z_l = 0.3$ & 17.5 & 1.92 &  7.662 &  7.673 &  7.655 & 1.00 & -0.0041 & -0.0583 \\
~           & 25.0 & 2.11 &  8.421 &  8.431 &  8.410 & 1.34 & -0.0035 & -0.0555 \\
$z_s = 0.8$ & 32.5 & 2.28 &  9.102 &  9.111 &  9.087 & 1.70 & -0.0031 & -0.0533 \\
~           & 40.0 & 2.43 &  9.731 &  9.740 &  9.712 & 2.08 & -0.0029 & -0.0515 \\
\hline
~           &  5.0 & 1.34 &  5.374 &  5.389 &  5.377 & 0.34 & -0.0087 & -0.0696 \\
$z_l = 0.3$ &  9.0 & 1.49 &  5.973 &  5.986 &  5.971 & 0.48 & -0.0066 & -0.0649 \\
~           & 13.0 & 1.61 &  6.455 &  6.467 &  6.450 & 0.60 & -0.0054 & -0.0618 \\
$z_s = 1.5$ & 17.0 & 1.72 &  6.878 &  6.889 &  6.870 & 0.72 & -0.0047 & -0.0595 \\
~           & 21.0 & 1.82 &  7.264 &  7.274 &  7.253 & 0.86 & -0.0042 & -0.0577 \\
\hline
~           & 10.0 & 2.05 &  8.202 &  8.214 &  8.199 & 1.76 & -0.0047 & -0.0591 \\
$z_l = 0.6$ & 17.5 & 2.37 &  9.472 &  9.485 &  9.464 & 2.72 & -0.0039 & -0.0551 \\
~           & 25.0 & 2.64 & 10.556 & 10.568 & 10.544 & 3.76 & -0.0036 & -0.0525 \\
$z_s = 0.8$ & 32.5 & 2.88 & 11.533 & 11.547 & 11.518 & 4.92 & -0.0035 & -0.0505 \\
~           & 40.0 & 3.11 & 12.440 & 12.455 & 12.422 & 6.18 & -0.0036 & -0.0489 \\
\hline
~           &  5.0 & 1.30 &  5.206 &  5.221 &  5.205 & 0.72 & -0.0091 & -0.0680 \\
$z_l = 1.0$ &  9.0 & 1.48 &  5.937 &  5.950 &  5.930 & 1.06 & -0.0067 & -0.0622 \\
~           & 13.0 & 1.64 &  6.545 &  6.557 &  6.533 & 1.44 & -0.0055 & -0.0584 \\
$z_s = 1.5$ & 17.0 & 1.77 &  7.087 &  7.099 &  7.071 & 1.82 & -0.0048 & -0.0555 \\
~           & 21.0 & 1.90 &  7.587 &  7.598 &  7.567 & 2.24 & -0.0044 & -0.0532 \\
\hline
\end{tabular}

\caption{The concentration parameters and relative errors in the
virial mass for a set of hypothetical lensing systems assuming that
the true system has a density profile $\rho_1$ with a scale radius
of $r_s=0.25$ Mpc and $\tau=2c$. We give the true $c$ value, and
values obtained by thin lens approximations using a density profile
$\rho_1$  ($c_1$) and an NFW profile $\rho_0$ ($c_0$). The virial
radius quoted, $r_{\mbox{200}}$, is the one calculated using the
true $c$ value. The true mass, $M^{\mbox{\scriptsize 200}}$ is given
in units of $10^{15}$ solar masses.\label{t2c:table} }
\end{center}
\end{table}


\clearpage

\begin{table} \begin{center}\begin{tabular}{rrcccccrr}
\hline $\tau=10c$ & & & & & & & \\
\hline System &  $\theta_E$ & $r_{200}$ (Mpc) &
$c$ & $c_1$ & $c_0$ & $M^{\mbox{\scriptsize 200}}~(10^{15} M_\odot)$ & $1-M^{\mbox{\scriptsize 200}}_1/M^{\mbox{\scriptsize 200}}$ & $1-M^{\mbox{\scriptsize 200}}_0/M^{\mbox{\scriptsize 200}}$ \\
\hline ~&~&~&~&~&~&~&~&~\\
~           & 10.0 & 2.05 &  8.199 &  8.205 &  8.204 & 1.18 & -0.0022 & -0.0045 \\
$z_l = 0.2$ & 17.5 & 2.29 &  9.149 &  9.155 &  9.154 & 1.64 & -0.0020 & -0.0042 \\
~           & 25.0 & 2.48 &  9.934 &  9.941 &  9.940 & 2.08 & -0.0020 & -0.0041 \\
$z_s = 0.4$ & 32.5 & 2.66 & 10.633 & 10.640 & 10.639 & 2.56 & -0.0020 & -0.0041 \\
~           & 40.0 & 2.82 & 11.274 & 11.282 & 11.281 & 3.06 & -0.0021 & -0.0041 \\
\hline
~           & 10.0 & 1.69 &  6.753 &  6.756 &  6.755 & 0.74 & -0.0014 & -0.0037 \\
$z_l = 0.3$ & 17.5 & 1.91 &  7.652 &  7.656 &  7.655 & 1.06 & -0.0013 & -0.0034 \\
~           & 25.0 & 2.10 &  8.408 &  8.411 &  8.410 & 1.42 & -0.0013 & -0.0034 \\
$z_s = 0.8$ & 32.5 & 2.27 &  9.084 &  9.089 &  9.087 & 1.78 & -0.0014 & -0.0033 \\
~           & 40.0 & 2.43 &  9.710 &  9.714 &  9.712 & 2.08 & -0.0014 & -0.0033 \\
\hline
~           &  5.0 & 1.34 &  5.375 &  5.378 &  5.377 & 0.36 & -0.0015 & -0.0038 \\
$z_l = 0.3$ &  9.0 & 1.49 &  5.970 &  5.972 &  5.971 & 0.50 & -0.0012 & -0.0035 \\
~           & 13.0 & 1.61 &  6.449 &  6.451 &  6.450 & 0.64 & -0.0011 & -0.0033 \\
$z_s = 1.5$ & 17.0 & 1.72 &  6.869 &  6.871 &  6.870 & 0.76 & -0.0011 & -0.0032 \\
~           & 21.0 & 1.81 &  7.252 &  7.254 &  7.253 & 0.90 & -0.0011 & -0.0031 \\
\hline
~           & 10.0 & 2.05 &  8.193 &  8.200 &  8.199 & 1.86 & -0.0023 & -0.0045 \\
$z_l = 0.6$ & 17.5 & 2.36 &  9.458 &  9.466 &  9.464 & 2.86 & -0.0023 & -0.0044 \\
~           & 25.0 & 2.63 & 10.536 & 10.545 & 10.543 & 3.94 & -0.0025 & -0.0044 \\
$z_s = 0.8$ & 32.5 & 2.88 & 11.510 & 11.520 & 11.518 & 5.14 & -0.0027 & -0.0046 \\
~           & 40.0 & 3.10 & 12.412 & 12.424 & 12.422 & 6.44 & -0.0030 & -0.0047 \\
\hline
~           &  5.0 & 1.30 &  5.203 &  5.206 &  5.205 & 0.76 & -0.0020 & -0.0042 \\
$z_l = 1.0$ &  9.0 & 1.48 &  5.928 &  5.932 &  5.930 & 1.12 & -0.0018 & -0.0039 \\
~           & 13.0 & 1.63 &  6.531 &  6.535 &  6.533 & 1.50 & -0.0018 & -0.0037 \\
$z_s = 1.5$ & 17.0 & 1.77 &  7.069 &  7.073 &  7.071 & 1.92 & -0.0018 & -0.0037 \\
~           & 21.0 & 1.89 &  7.565 &  7.569 &  7.567 & 2.34 & -0.0019 & -0.0037 \\
\hline

\end{tabular} \caption{The concentration parameters and relative
errors in the virial mass for a set of hypothetical lensing systems
assuming that the true system has a density profile $\rho_1$ with a
scale radius of $r_s=0.25$ Mpc and $\tau=10c$. We give the true $c$
value, and values obtained by thin lens approximations using a
density profile $\rho_1$  ($c_1$) and an NFW profile $\rho_0$
($c_0$). The virial radius quoted, $r_{\mbox{200}}$, is the noe
calculated using the true $c$ value.  The true mass,
$M^{\mbox{\scriptsize 200}}$ is given in units of $10^{15}$ solar
masses.\label{t10c:table} }
\end{center}
\end{table}

\begin{table} \begin{center}\begin{tabular}{rrcccccr}
\hline $r_s = 0.25$ Mpc & & & & & & & \\
\hline System &  $\theta_E$ & $r_{200}^{\mbox{\scriptsize soft}}$
(Mpc) &
$c_{\mbox{\scriptsize soft}}$ & $c_{\mbox{\scriptsize hard}}$ & $M^{\mbox{\scriptsize 200}}_{\mbox{\scriptsize soft}}~(10^{15} M_\odot)$ & $1-M^{\mbox{\scriptsize 200}}_{\mbox{\scriptsize hard}}/M^{\mbox{\scriptsize 200}}_{\mbox{\scriptsize soft}}$ \\
\hline ~&~&~&~&~&~&~&~\\
~           & 10.0 & 2.05 &  8.201 &  8.209 & 1.15 & -0.0308 \\
$z_l = 0.2$ & 17.5 & 2.29 &  9.153 &  9.161 & 1.59 & -0.0296 \\
~           & 25.0 & 2.48 &  9.939 &  9.947 & 2.04 & -0.0287 \\
$z_s = 0.4$ & 32.5 & 2.66 & 10.639 & 10.647 & 2.51 & -0.0281 \\
~           & 40.0 & 2.82 & 11.282 & 11.290 & 2.99 & -0.0276 \\
\hline
~           & 10.0 & 1.69 &  6.756 &  6.766 & 0.71 & -0.0340 \\
$z_l = 0.3$ & 17.5 & 1.91 &  7.658 &  7.668 & 1.04 & -0.0324 \\
~           & 25.0 & 2.10 &  8.415 &  8.425 & 1.38 & -0.0314 \\
$z_s = 0.8$ & 32.5 & 2.27 &  9.093 &  9.104 & 1.74 & -0.0306 \\
~           & 40.0 & 2.43 &  9.720 &  9.731 & 2.13 & -0.0300 \\
\hline
~           &  5.0 & 1.34 &  5.376 &  5.388 & 0.36 & -0.0380 \\
$z_l = 0.3$ &  9.0 & 1.49 &  5.972 &  5.984 & 0.49 & -0.0363 \\
~           & 13.0 & 1.61 &  6.453 &  6.464 & 0.62 & -0.0351 \\
$z_s = 1.5$ & 17.0 & 1.72 &  6.874 &  6.886 & 0.75 & -0.0343 \\
~           & 21.0 & 1.81 &  7.258 &  7.270 & 0.88 & -0.0336 \\
\hline
~           & 10.0 & 2.05 &  8.198 &  8.207 & 1.81 & -0.0313 \\
$z_l = 0.6$ & 17.5 & 2.37 &  9.466 &  9.475 & 2.79 & -0.0298 \\
~           & 25.0 & 2.64 & 10.546 & 10.556 & 3.86 & -0.0288 \\
$z_s = 0.8$ & 32.5 & 2.88 & 11.521 & 11.532 & 5.04 & -0.0280 \\
~           & 40.0 & 3.11 & 12.426 & 12.437 & 6.33 & -0.0274 \\
\hline
~           &  5.0 & 1.30 &  5.205 &  5.219 & 0.74 & -0.0396 \\
$z_l = 1.0$ &  9.0 & 1.48 &  5.933 &  5.947 & 1.10 & -0.0374 \\
~           & 13.0 & 1.63 &  6.538 &  6.552 & 1.47 & -0.0361 \\
$z_s = 1.5$ & 17.0 & 1.77 &  7.078 &  7.093 & 1.87 & -0.0351 \\
~           & 21.0 & 1.89 &  7.576 &  7.591 & 2.30 & -0.0344 \\
\hline
\end{tabular} \caption{The concentration parameters and relative
error in the virial mass for a set of possible lensing systems
modeled with soft truncation versus hard truncation.  For soft
truncation, the density profile is $\rho_1$ with $\tau=3c$. For hard
truncation, the density profile is $\rho_0$ within the virial
radius, and zero outside of the virial radius. The corresponding
values of $c$ are both found by integration of the geodesic
equations. \label{comparison:table} }
\end{center}
\end{table}


\clearpage
\begin{table} \begin{center}\begin{tabular}{rccc}
\hline System &  $z_l$ & $z_s$ & $\theta_E$ (arc sec)  \\
\hline ~&~&~&~ \\

RXJ1347-1145 & 0.45 & 0.80 & 35 \\
MS2137+23 & 0.313 & 1.6 & 16 \\
RCS2319+00A & 0.9 & 3.86 & 12 \\

\hline

\end{tabular} \caption{\label{systems:table}
Three lensing systems analyzed in Figs.~\ref{t3c:fig}, \ref{t2c:fig}
and \ref{j:fig}. System RXJ1347-1145 is from Sahu et al. (1998),
system MS2137+23 is from Gavazzi et al. (2003) and system RCS2319+00
is from Gilbank et al. (2008). }
\end{center}
\end{table}


\clearpage

\clearpage

\end{document}